*Article*

# Using Perspective-n-Point Algorithms for a Local Positioning System Based on LEDs and a QADA Receiver

Elena Aparicio-Esteve, Jesús Ureña *, Álvaro Hernández, Daniel Pizarro and David Moltó

Electronics Department, University of Alcalá, 28801 Madrid, Spain; elena.aparicio@uah.es (E.A.-E.); alvaro.hernandez@uah.es (Á.H.); daniel.pizarro@uah.es (D.P.); david.molto@edu.uah.es (D.M.)
* Correspondence: jesus.urena@uah.es

**Abstract:** The research interest on location-based services has increased during the last years ever since 3D centimetre accuracy inside intelligent environments could be confronted with. This work proposes an indoor local positioning system based on LED lighting, transmitted from a set of beacons to a receiver. The receiver is based on a quadrant photodiode angular diversity aperture (QADA) plus an aperture placed over it. This configuration can be modelled as a perspective camera, where the image position of the transmitters can be used to recover the receiver's 3D pose. This process is known as the perspective-n-point (PnP) problem, which is well known in computer vision and photogrammetry. This work investigates the use of different state-of-the-art PnP algorithms to localize the receiver in a large space of $2 \times 2\,\text{m}^2$ based on four co-planar transmitters and with a distance from transmitters to receiver up to 3.4 m. Encoding techniques are used to permit the simultaneous emission of all the transmitted signals and their processing in the receiver. In addition, correlation techniques (match filtering) are used to determine the image points projected from each emitter on the QADA. This work uses Monte Carlo simulations to characterize the absolute errors for a grid of test points under noisy measurements, as well as the robustness of the system when varying the 3D location of one transmitter. The IPPE algorithm obtained the best performance in this configuration. The proposal has also been experimentally evaluated in a real setup. The estimation of the receiver's position at three particular points for roll angles of the receiver of $\gamma = \{0°, 120°, 210° \text{ and } 300°\}$ using the IPPE algorithm achieves average absolute errors and standard deviations of 4.33 cm, 3.51 cm and 28.90 cm; and 1.84 cm, 1.17 cm and 19.80 cm in the coordinates $x$, $y$ and $z$, respectively. These positioning results are in line with those obtained in previous work using triangulation techniques but with the addition that the complete pose of the receiver ($x$, $y$, $z$, $\alpha$, $\beta$, $\gamma$) is obtained in this proposal.

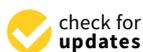

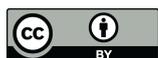

## 1. Introduction

Ambient Intelligence is spreading worldwide with the development of multiple applications and services focused on users and businesses. One of them is the so-called Location-based Services (LBS), where a set of applications is offered to users depending on their position inside the intelligent space or environment [1]. These services pose a new challenge consisting in the accurate determination of people, robots or any other device not only located outdoors where global navigation satellite systems (GNSS) provide a feasible solution but also located indoors where no global solution has emerged yet in the market.

In this context, an indoor local positioning system (LPS) is often required and designed to estimate the pose of people or devices in environments where GNSS solutions have attenuation/coverage problems and multi-path effects that limit their performance [2]. For that purpose, different sensory technologies have also been applied in previous work (acoustic [3,4], infrared [5,6], radio-frequency [7,8], etc.). In particular, the use of optical signals (infrared or visible light) [9] has spread significantly due to their low cost, long lifetime and their presence in most of today's infrastructures [10].

Both Infrared and Visible Light Local Positioning Systems (IRLPS and VLPS) typically employ LEDs as transmitters and imaging sensors or photodetectors as receivers. Those based on imaging sensors use cameras in commercial smartphones [11] or CMOS cameras [12]. If the receptor is a photodetector, it often consists of an array of photodiodes [13]. The most common photoreceivers used in visible and near infrared light detection are position-sensitive detectors (PSD) in combination with a lens [14,15] and quadrant photodiode angular diversity aperture (QADA) with an aperture lens [16,17]. In these sensors, the perspective camera model (also known as the pin-hole camera) is an accurate geometric model to relate the beacon's 3D position with the sensor's 2D measurement. This model is widely used in photogrammetry and computer vision to solve problems that involve the 3D geometry of multiple views [18]. One such problem is known as the Perspective-n-Point (PnP) problem. Given a collection of 3D and 2D point correspondences between the scene and the image, the objective in PnP is to recover the camera pose (i.e., 6 degrees-of-freedom). Since the geometry of the LPS beacons is commonly known, the problem of finding the absolute orientation and position of the LPS receiver (i.e., the camera) can be exactly cast as an instance of the PnP problem.

This work focuses on the design and the experimental validation of an infrared positioning system based on a set of $n = 4$ transmitting LEDs (beacons) located at known coplanar positions. Thus, we deal with a non-redundant amount of beacons in the Planar-PnP case. The receiver is based on a QADA sensor with a square aperture. This system is intended for long-distance measurements (up to 4.5 m), and it is based on the reception and further processing of different encoded signals, which are simultaneously emitted by the aforementioned LEDs [19]. The system processes the received signals by applying a matched filter to obtain the maximum correlation values for each transmission. Based on these values, the 2D projections from the beacons on the QADA photodetector are estimated [19]. After obtaining these image points, we apply recent PnP solutions that handle Planar-PnP, such as EPnP, IPPE and RPnP, to obtain the 6-DOF pose of the receiver. A real prototype has been assembled to successfully validate the approach in an experimental setup, where the distance between the plane of the beacons in the ceiling and the receiver in the floor is 3.4 m (room's height). It is worth noting that these PnP algorithms estimate the complete pose of the receiver ($x$, $y$, $z$, $\alpha$, $\beta$ and $\gamma$).

Previous investigations have already used triangulation algorithms in a similar scenario to estimate the 3D position ($x$, $y$ and $z$) and the rotation of the mobile in the $z$-axis [16,20]. Nevertheless, this work presents a 3D IRLPS for which its main novelty is the involvement of recent PnP solutions and a low number of beacons to estimate the entire pose of a mobile robot ($x$, $y$, $z$, $\alpha$, $\beta$ and $\gamma$) in cases with long distances between the emitters and the receiver where a low SNR is expected. With regard to the system characterization, the different measurements and analysis carried out hereinafter assume that the transmitters are placed in the ceiling (not tilted nor rotated), whereas the receiver can move in a certain plane (i.e., the floor) with any rotation $\gamma$ around the $z$ axis. Further improvements in terms of more degrees of freedom can also be derived. In particular, in the current configuration, the transmitters are located in the same plane; however, this constraint can be released since both EPnP and RPnP algorithms deal with non-planar cases, although at the cost of increasing the number of transmitters. Similarly, another possible enhancement is related to the plane where the receiver is located. The proposed 3D IRLPS can be extended to obtain the pose of a mobile receiver ($x$, $y$, $z$, $\alpha$, $\beta$ and $\gamma$) by using planar or non-planar algorithms but at the expense of decreasing the accuracy.

The rest of the manuscript is organized as follows: Section 2 presents related works found in the literature; Section 3 provides a general overview of the proposed system; Section 4 describes the applied PnP algorithms; Section 5 details the performance of the proposal; Section 6 shows some experimental results obtained in some selected points from the coverage area; and, finally, conclusions are discussed in Section 7.

## 2. Related Work

Practical systems involving indoor positioning usually rely on data acquired from mobile sensors (e.g., inertial sensors), which are prone to fluctuations caused by the environment and accumulate errors. The performance of these systems depends on the correction and ease of removal of these errors [21]. Numerous approaches have attempted to solve this problem by using secondary sensors, e.g., ultrasonic-based or laser-based ranging sensors [3], blinking LEDs [5], magnetic sensors, etc. Another option is the use of landmarks, which can be artificial (developed and conveniently placed for a particular application) or natural (detected as particular features of the environment). For instance, Visual Simultaneous Localization and Mapping (vSLAM) is a technology introduced in [22] for creating a visual landmark map. They use an object recognition algorithm on a camera image captured by a mobile device. This technology does not require the installation of additional infrastructure or auxiliary devices, but it has to deal with large landmark databases and large computational times.

As in vision-based landmark systems in which the camera pose is obtained from a collection of 3D and 2D point correspondences between the scene (landmarks) and the image, this investigation deals with a similar problem of finding the absolute orientation and position of the LPS receiver (i.e., the QADA) and knowing the correspondences between the 3D LED positions and the 2D surface of the receiver. We are particularly interested in the analysis of the performance of previously developed PnP algorithms for this type of application.

The PnP problem has been thoroughly investigated in computer vision and mathematics. The minimal case occurs when $n = 3$, namely P3P, and yields four possible solutions [23]. Closed-form solutions for the P3P problem are known for more than a century. For $n \leq 4$, the problem has a unique solution in general non-degenerate configurations. When the $n$ points lie in a plane, the problem is known as Planar-PnP. Existing PnP solutions can be divided into iterative and non-iterative methods.

Non-iterative PnP methods use convex relaxations that admit a closed-form solution [24,25] or are based on finding the roots of polynomial equations [23,26]. Some representative non-iterative methods from the state-of-the-art include Efficient Perspective-n-Point Camera Pose Estimation (EPnP) [24], Robust Non-Iterative Solution of Perspective-n-Point (RPnP) [26] and Infinitesimal Plane-Based Pose Estimation (IPPE) [27], which exclusively solves Planar-PnP. EPnP is usually sub-optimal for $n < 5$, and it requires a special configuration to work in the planar or quasi-planar case [26]. RPnP is accurate for both planar and non-planar cases.

Finally, iterative PnP methods are often more accurate than non-iterative methods. They are based on optimizing a non-convex Maximum-Likelihood reprojection cost with iterative descent methods. These methods have high computational cost and require initialization near the optimum point, usually with a non-iterative method [24,26].

## 3. Global Proposed System Description

The proposed infrared positioning system is based on $n = 4$ LED emitters or beacons placed at known locations in the ceiling plane so that they cover a certain volume where the receiver can detect all the transmissions in order to obtain its own location. If a larger system is required, the proposed system can be easily extended by increasing the number of LEDs deployed in the environment. Note that each LED has its own different code to avoid interference and facilitate their independent operation. Every four LEDs can be treated as a group by using the method presented in this work, adapted to their particular coverage area.

Figure 1 shows a general scheme of the proposed IRLPS. In this proposal, we consider three independent coordinate systems: $x$, $y$ and $z$ denote the Cartesian coordinates in the global coordinate system, with the origin at the corner of the room. The camera coordinates are denoted by $x_{cam}$, $y_{cam}$ and $z_{cam}$ with the origin located at the center of the square aperture in the receiver. Finally, $x_r$ and $y_r$ represent the local 2D coordinates of the photoreceiver QADA in which its origin is placed at the centre of the QADA. The receiver's complete pose is denoted as $x$, $y$, $z$, $\alpha$, $\beta$ and $\gamma$, where $\gamma$ is the rotation angle around the $z$ axis of the receiver. Note that since the receiver is placed on top of a mobile robot and the $x - y$ plane corresponds to the floor plane, then $\alpha = 0$ and $\beta = 0$.

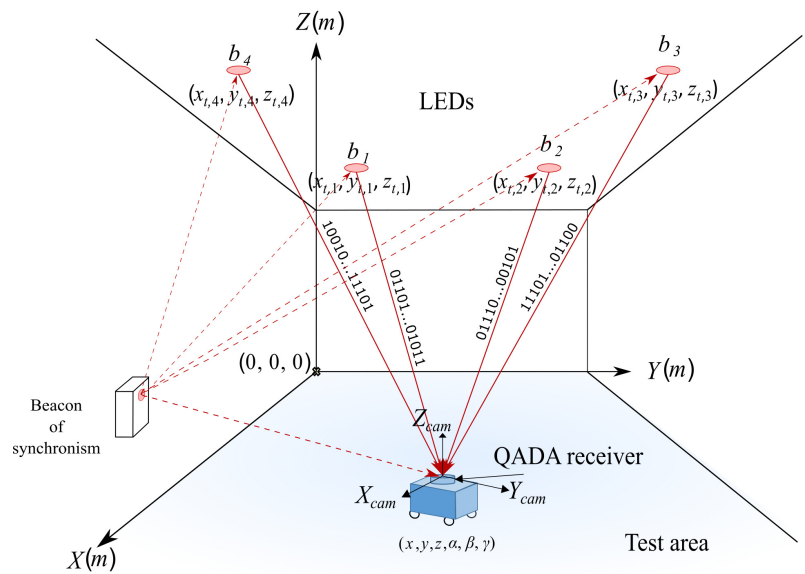

**Figure 1.** Global overview of the proposed system.

It is worth noting that this is a synchronized approach where the receiver starts to acquire at the same time as the transmitters emit and when the synchronization pulse, emitted by the synchronizing beacon, is received by both emitters and receiver. Since the later positioning algorithm requires the identification of the different transmissions coming from the four beacons at the receiver, a Code-Division Multiple Access (CDMA) technique is applied, so every LED $i$ transmits an unequivocal code $c_i$. The receiver can recognise these codes by applying the corresponding matched filters.

The reception system is based on (1) a QADA circular photoreceptor QP50-6-18u-TO8 [28] in addition to an aperture placed over it in such a manner that the incident signal passes through the aperture and irradiates part of its surface; (2) a filtering stage that reduces any undesired signal; (3) a synchronism detector; and (4) an STM32F469I Discovery acquisition system [29]. As it can be observed in Figures 1 and 2, the proposed system can be considered as a pin-hole model with regard to the incident light on the QADA receiver, where the Line-of-Sight (LoS) measurements can be used to obtain the pose of the receiver. The incident light coming from the transmitters generates four different currents $i_j(t)$ where each current is for each quadrant $j = \{1, 2, 3, 4\}$. These currents are combined in the receiver module in order to provide the difference of voltages in the X axis ($V_{lr}$), in the Y axis ($V_{bt}$) and for the sum of all signals ($V_{sum}$).

At the reception stage, in order to identify each transmitter, the system simultaneously correlates the acquired signals $V_{sum}$, $V_{bt}$ and $V_{lr}$ with the transmitted codes $c_i$, while rejecting other external signals (incoming sunlight, ambient light and noise, etc.) provided that the receiver circuit is not saturated. We refer to the signal correlations of the signals $V_{sum}$, $V_{lr}$ and $V_{bt}$ as $s_i[n]$, $r_i[n]$ and $t_i[n]$, respectively. The maximum values of the correlated signals are combined as follows.

$$p_x = \frac{max(r_i[n])}{max(s_i[n])} \quad (1)$$

$$p_y = \frac{max(t_i[n])}{max(s_i[n])} \quad (2)$$

It is crucial to acquire the complete sequence $c_i$ without interferences in order to accurately estimate the ratios $p_x$ and $p_y$. These are used to estimate the position of the image point ($x_r$ and $y_r$) for every transmitter in the QADA photodiode according to the following:

$$\begin{bmatrix} x_r \\ y_r \end{bmatrix} = \frac{-l}{2} \cdot \lambda \cdot \begin{bmatrix} p_x + \delta \cdot p_y \\ -\delta \cdot p_x + p_y \end{bmatrix} + \begin{bmatrix} x_c \\ y_c \end{bmatrix} \quad (3)$$

where the aperture misalignment $\delta$, the central point ($x_c$ and $y_c$), the aperture length $l$ and the ratio between the expected focal length $h_{ap}$ and the actual focal length $h'_{ap}$ where $\lambda = \frac{h'_{ap}}{h_{ap}}$ have all been considered [20]. The image points in Equation (3) are geometrically related with the transmitter's 3D coordinates as follows:

$$x_r = h_{ap} \cdot \frac{x_{cam}}{z_{cam}}$$
$$y_r = h_{ap} \cdot \frac{y_{cam}}{z_{cam}} \quad (4)$$

where $x_{cam}$, $y_{cam}$ and $z_{cam}$ are the transmitter position in camera coordinates obtained as follows:

$$\begin{bmatrix} x_{cam} \\ y_{cam} \\ z_{cam} \end{bmatrix} = [R|t] \cdot \begin{bmatrix} x_t \\ y_t \\ z_t \end{bmatrix} \quad (5)$$

where $[R|t]$ concatenates the rotation matrix $R = R(\alpha, \beta, \gamma)$, and the translation matrix $t = (x, y, z)^\top$ defined by the receiver's pose and $x_t$, $y_t$ and $z_t$ is the transmitter's position in the global coordinate system. Equation (4) corresponds to the pin-hole projection model with the following intrinsic parameters.

$$K = \begin{bmatrix} h_{ap} & 0 & 0 \\ 0 & h_{ap} & 0 \\ 0 & 0 & 1 \end{bmatrix} \quad (6)$$

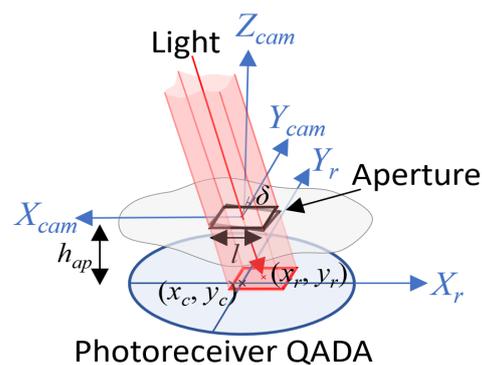

**Figure 2.** Representation of the geometry of the QADA sensor and the aperture over it.

Once the image points for each transmitter $i$ are obtained, different PnP solutions can be directly applied to estimate the pose of the receiver in the proposed scenario. The input of these algorithms are the image points $(x_r, y_r)_i$, the focal length $h_{ap}$ which defines the intrinsics' matrix and the beacons' position $(x_t, y_t, z_t)_i$. The output of the PnP solutions is the complete 3-D pose of the receiver ($x$, $y$, $z$, $\alpha$, $\beta$ and $\gamma$). More detailed information about the presented positioning algorithm can be found in the pseudo-code presented in Algorithm 1 and in previous works [19,30,31].

**Algorithm 1** Positioning algorithm
―――――――――――――――――――――――――――――――――――――
1: noLEDs = 4
2: **for** $i = 1$ : noLEDs **do**
3: 　　Transmitter $i$ emits code $c_i$
4: **end for**
5: $[V_{sum}, V_{bt}, V_{lr}]$ = SignalAcquisition ($c_i$)
6: **for** $i = 1$ : noLEDs **do**
7: 　　$[s[n], t[n], r[n]]_i$ = Correlation ($[V_{sum}, V_{bt}, V_{lr}], c_i$)
8: 　　$[p_x, p_y]_i = \frac{[max(r[n]), max(t[n])]_i}{max(s[n]_i)}$
9: 　　$\begin{bmatrix} x_r \\ y_r \end{bmatrix}_i = \frac{-l}{2} \cdot \lambda \cdot \begin{bmatrix} p_{x,i} + \delta \cdot p_{y_i} \\ -\delta \cdot p_{x,i} + p_{y_i} \end{bmatrix} + \begin{bmatrix} x_c \\ y_c \end{bmatrix}$
10: **end for**
11: $[x, y, z, \alpha, \beta, \gamma]$ = PnP Algorithms ($x_r, y_r, h_{ap}$, beacons' position)
―――――――――――――――――――――――――――――――――――――

## 4. PnP Algorithms

In this work we employ three well established state-of-the-art PnP methods (i.e., EPnP, RPnP and IPPE) to find the pose of the receiver from the image points of each transmitter.

In EPnP [24], the $n$ 3D points are obtained from four virtual control points performing a weighted sum. The algorithm makes estimates in the receiver coordinate system on the 3D coordinates of these control points as a weighted sum of the eigenvectors of a $12 \times 12$ matrix and solves a small system of quadratic equations to select the appropriate weights. The solution can be used in the initialization of the Gauss–Newton refinement scheme in order to improve accuracy.

In the RPnP solution [26], the PnP problem is expressed as a least squares polynomial function $F$ derived from the classic P3P polynomial equations [32]. The minima of $F$ are found as the roots of its derivative $F'$ by using the eigenvalue method. This method explores the properties of this function minima and their geometric relationship with the degenerate configurations and the Planar-PnP case.

The IPPE solution [27] exclusively solves the Planar-PnP problem. First, it estimates the coefficients of the homographic transformation between the image coordinates and the 3D plane that contains the transmitters. Second, it finds the pose parameters from the homographic coefficients. This is a decomposition algorithm that exploits redundancy in the homographic coefficients to find the pose in closed-form, which maximizes pose accuracy. This involves solving a local non-redundant 1st order PDE. IPPE is very optimized in terms of computational complexity, being suitable to run in real time in lightweight computing hardware.

Note that each algorithm is initially designed for certain conditions and particular cases; therefore, their behaviour will be different regarding the proposed optical system where a planar configuration with four transmitters is deployed ($n = 4$). It is worth mentioning that EPnP and RPnP handle both planar and non-planar configurations: EPnP is more suitable for $n \geq 4$ [24], whereas RPnP presents a robust and high accuracy estimation of the pose when $n < 5$ [26].

## 5. Simulated Results

Hereinafter, the EPnP, IPPE and RPnP algorithms are applied to estimate the pose of the receiver using simulated data. In the following simulations, the volume under analysis is $2 \times 2 \times 3.4 \, \text{m}^3$. There is a distance of 3.4 m between the ceiling where the transmitters are installed and the floor where the receiver is placed. This receiver is tested on a grid of points defined in the floor with an interval of 10 cm. The number of measurements per analysed point is 50, and a Gaussian noise has been added in the received signal in order to have a Signal-to-Noise Ratio (SNR) of 10 dB. Note that the rotations considered in the receiver throughout these analyses are limited to the Z axis, $\gamma = \{0°, 120°, 240°\}$.

The mean absolute errors for each coordinate and for the aforementioned EPnP, IPPE and RPnP algorithms are plotted in Figures 3–5 with $\gamma = 120°$. It can be observed how the errors in the grid of considered points are linked to the distance between the transmitters and the receiver, thus providing slightly higher errors in the corners of the proposed scenario. Similar results are found for $\gamma = \{0°, 240°\}$.

For clarity's sake, the Cumulative Distribution Function (CDF) of the absolute pose error is presented in Figure 6 for the EPnP, IPPE and RPnP algorithms with $\gamma = \{0°, 120°, 240°\}$. The errors in the estimation of the 2D coordinates ($x$ and $y$) are below 10 cm for the IPPE and EPnP algorithms and below 20 cm for the RPnP algorithm. In addition, the absolute error related to coordinate $z$ is 3 cm, 7 cm and 30 cm for the IPPE, RPnP and EPnP algorithms, respectively. On the other hand, the errors in the estimation of the rotation angles $\alpha$ and $\beta$ are below 1.5° for the IPPE and EPnP algorithms, whereas the errors are below 4° for RPnP. Finally, the absolute errors in the estimation of $\gamma$ are 0.2°, 0.4° and 10° for IPPE, RPnP and EPnP, respectively.

In addition to the previous analysis, a particular CDF for nine points in the same scenario is presented in Figures 7–9 for the EPnP, IPPE and RPnP algorithms. These points are located in the first quadrant (see Table 1), as the rest of quadrants will have similar behaviours given the symmetries of the positioning system and considering the variations in all the rotation angles (0°–360°) with a step of 10°. The number of measurements per analysed point is 50, and the SNR is still 10 dB. The obtained absolute errors are similar to those in Figure 6. Note that the points in the corner of the area under analysis (points 3, 6, 7, 8 and 9) present slightly higher errors than those in the center of the analysed scenario (points 1, 2, 4 and 5). Moreover, although all the involved algorithms behave similarly, the IPPE presents better results for the proposed system in all the cases.

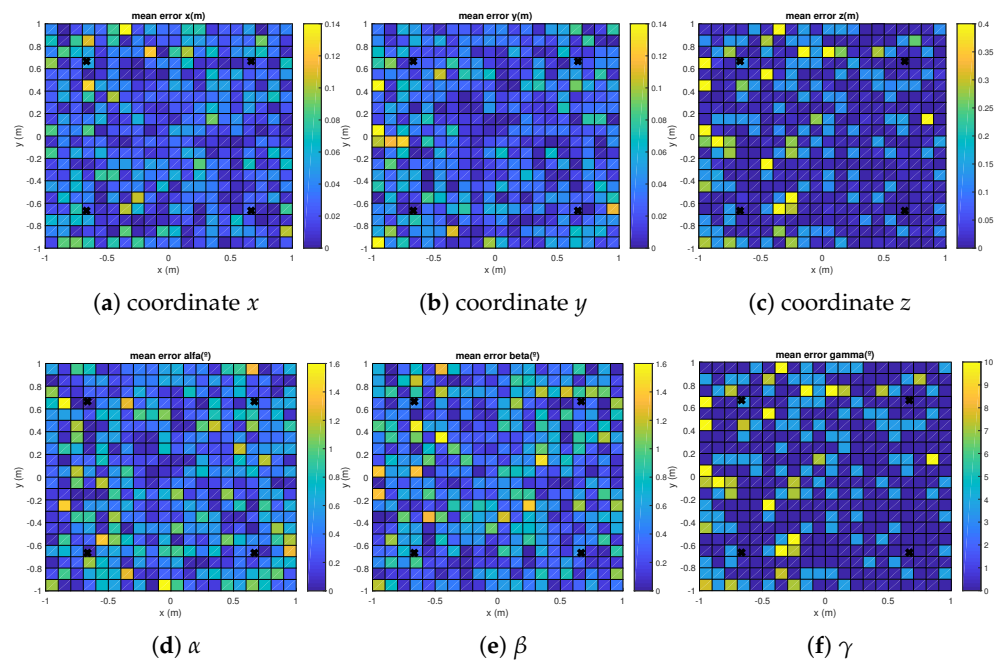

**Figure 3.** Mean absolute errors in the grid of considered points in the floor for the EPnP algorithm and a rotation in the Z axis $\gamma = 120°$.

Furthermore, in order to clearly select the algorithm with the best behaviour for the proposed system, we have analysed the robustness of the system in the estimation of the pose when varying the 3D location of one transmitter according to a Gaussian noise, $\{\sigma_x, \sigma_y, \sigma_z\} = 1$ cm. These simulations have been performed under the same conditions as before, with a volume of $2 \times 2 \times 3.4$ m$^3$ and a number of realization per point of 50. The CDF of the absolute pose error is presented in Figure 10 for the EPnP, IPPE and RPnP algorithms with $\gamma = \{0°, 120°, 240°\}$. All the considered algorithms present similar behaviours again, with a slightly better response from the IPPE algorithm. For IPPE, the absolute errors in the estimation of the 2D coordinates (*x* and *y*) are below 10 cm, and the absolute error related to coordinate *z* is less than 3 cm in 90% of cases. On the other hand, the absolute errors in the estimation of the rotation angles $\alpha$ and $\beta$ are below 1.5° for 90% of the cases, whereas $\gamma$ is below 0.2° also for 90% of the cases.

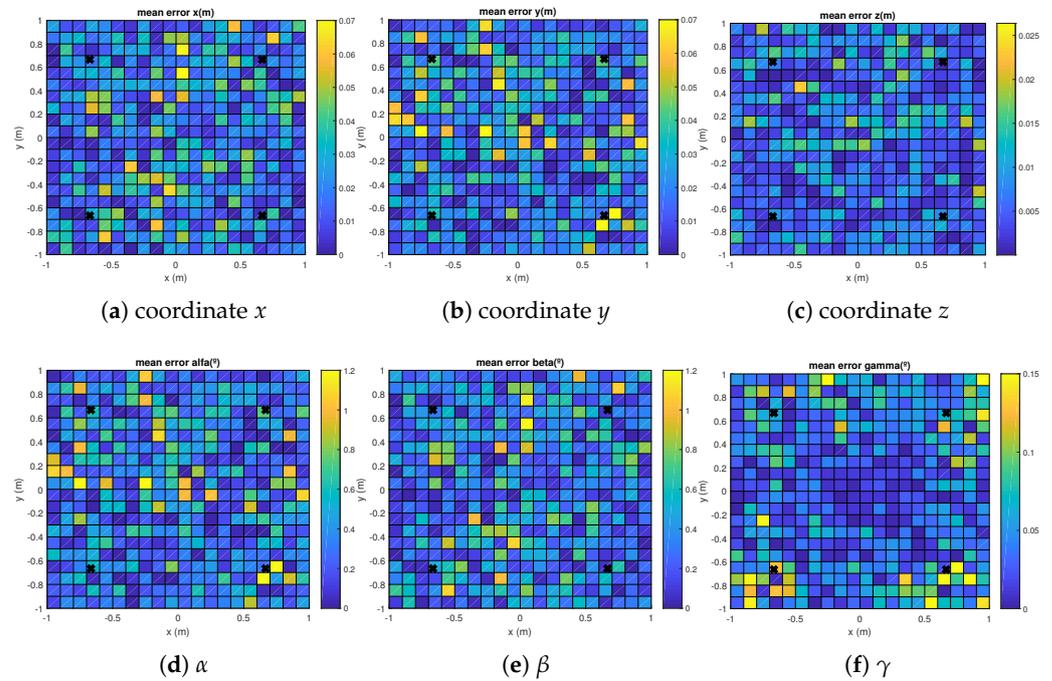

**Figure 4.** Mean absolute errors in the grid of considered points in the floor for the IPPE algorithm and a rotation in the Z axis $\gamma = 120°$.

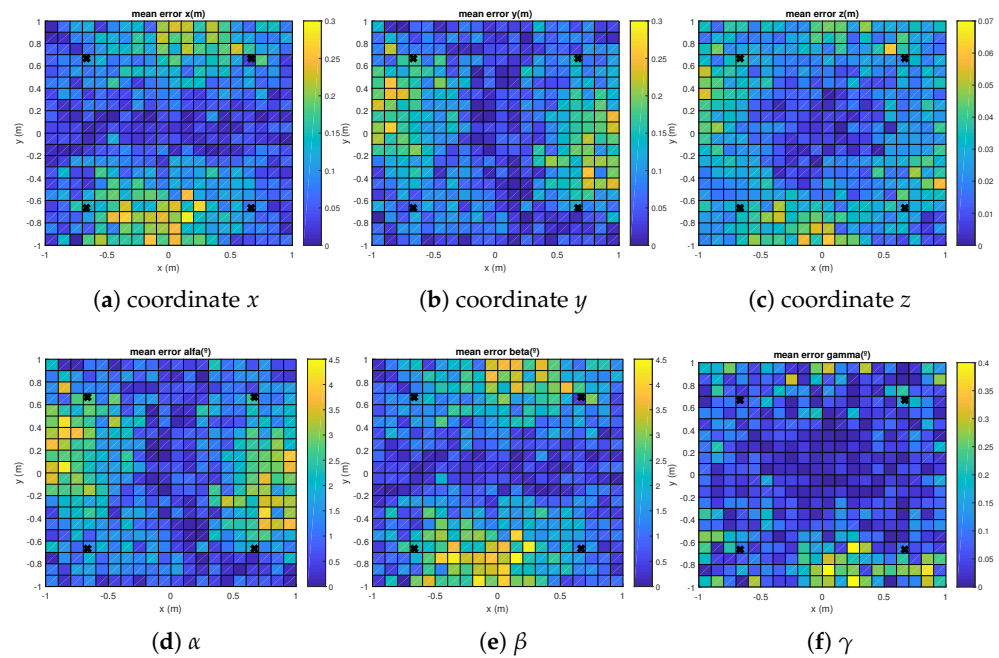

**Figure 5.** Mean absolute errors in the grid of considered points in the floor for the RPnP algorithm and a rotation in the Z axis $\gamma = 120°$.

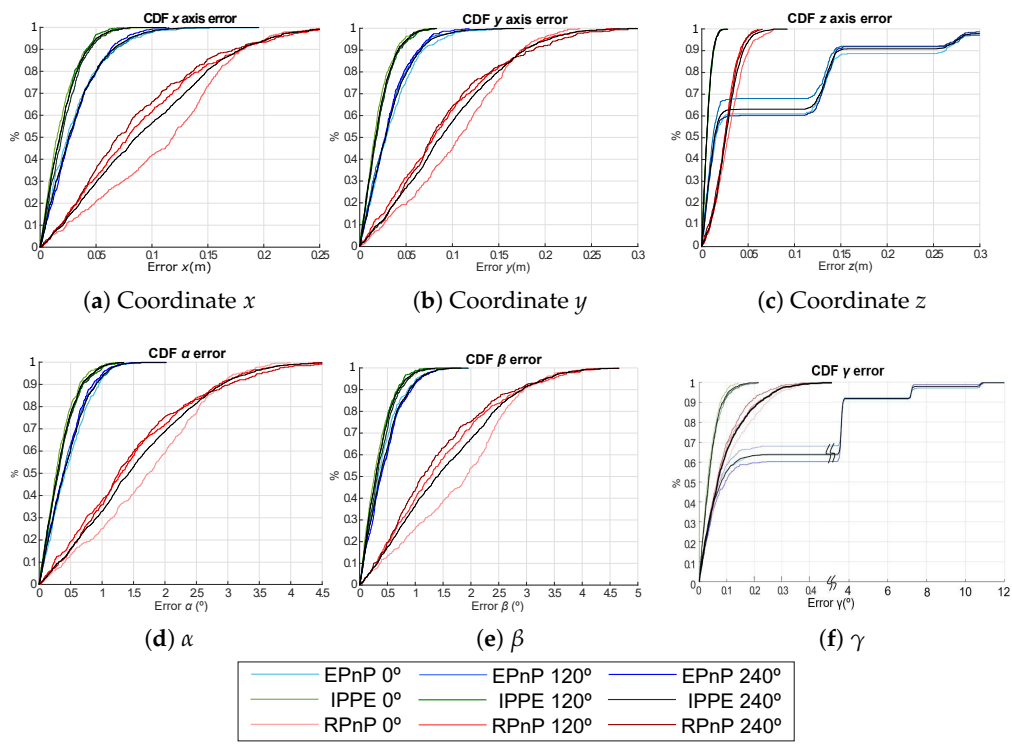

**Figure 6.** CDF of the absolute pose errors for EPnP, IPPE and RPnP according to the rotation in the $Z$ axis $\gamma = \{0°, 120°, 240°\}$.

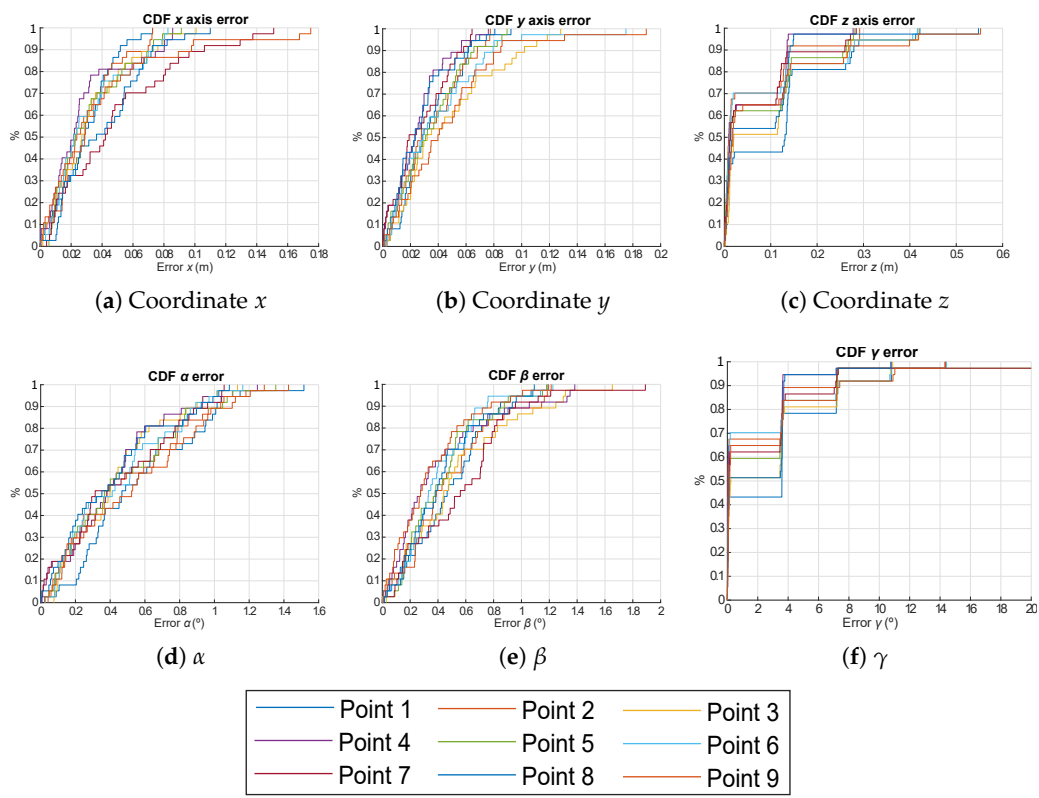

**Figure 7.** CDF of the absolute pose errors for the nine representative points defined in Table 1 for the case of using the EPnP algorithm.

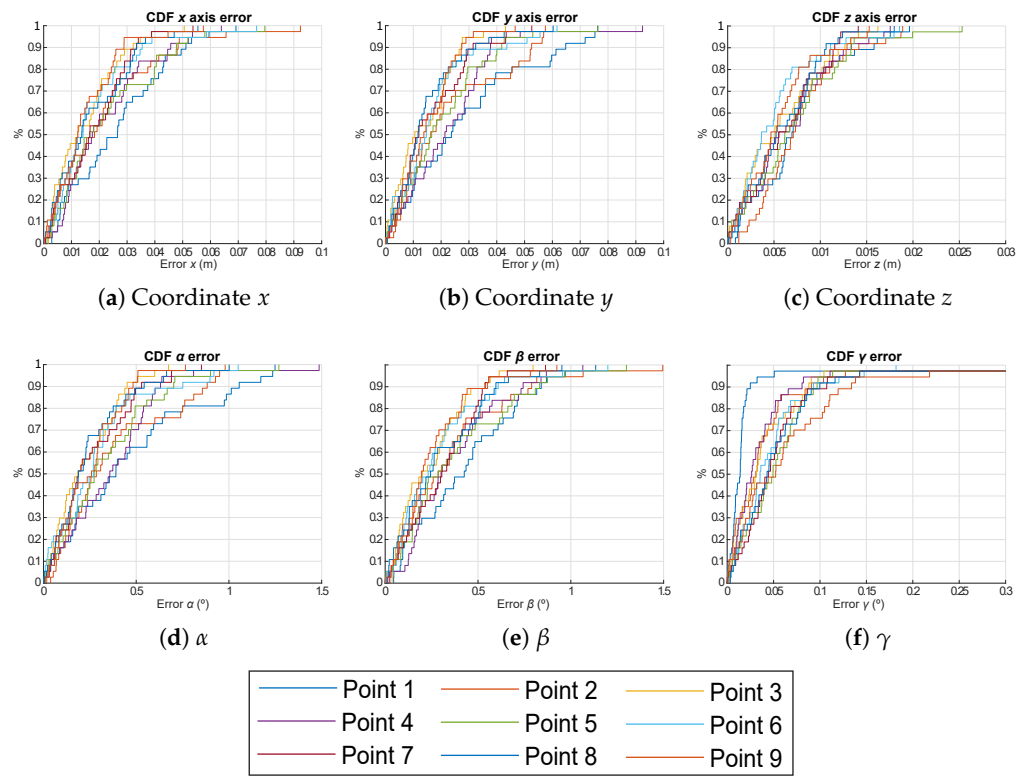

**Figure 8.** CDF of the absolute pose errors for the nine representative points defined in Table 1 for the case of using the IPPE algorithm.

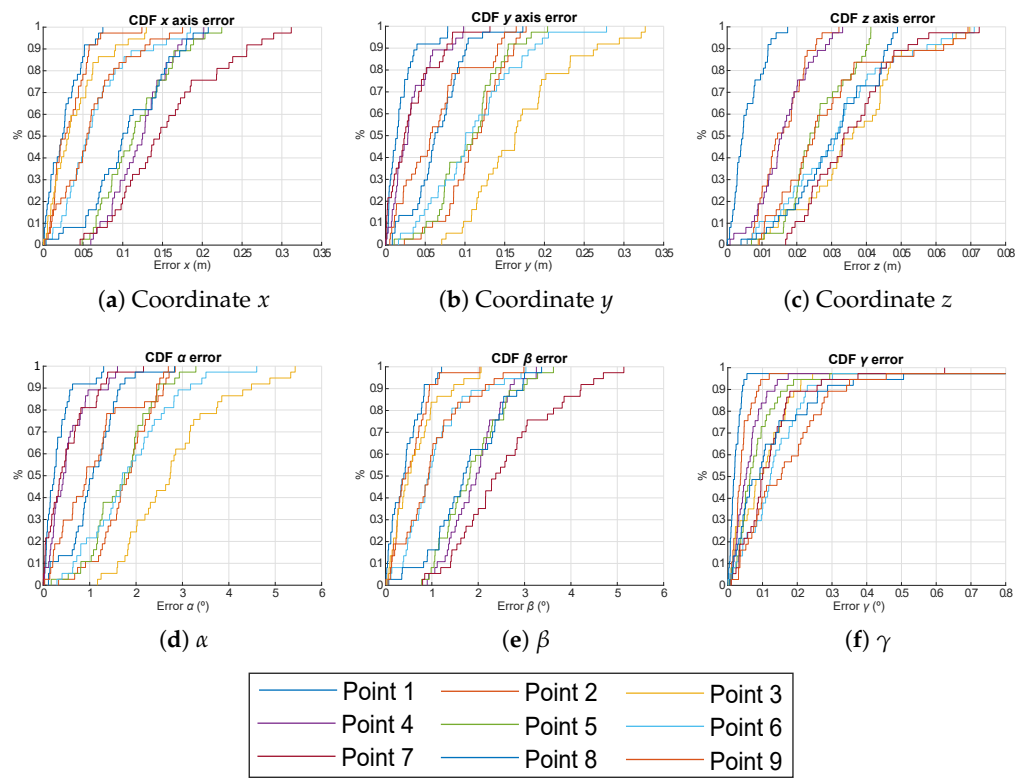

**Figure 9.** CDF of the absolute pose errors for the nine representative points defined in Table 1 for the case of using the RPnP algorithm.

**Table 1.** Coordinates of the nine representative points considered in the first quadrant for further analysis.

| Points | Coordinates (m) |
| --- | --- |
| Point 1 | (0, 0) |
| Point 2 | (0, 0.5) |
| Point 3 | (0, 1) |
| Point 4 | (0.5, 0) |
| Point 5 | (0.5, 0.5) |
| Point 6 | (0.5, 1) |
| Point 7 | (1, 0) |
| Point 8 | (1, 0.5) |
| Point 9 | (1, 1) |

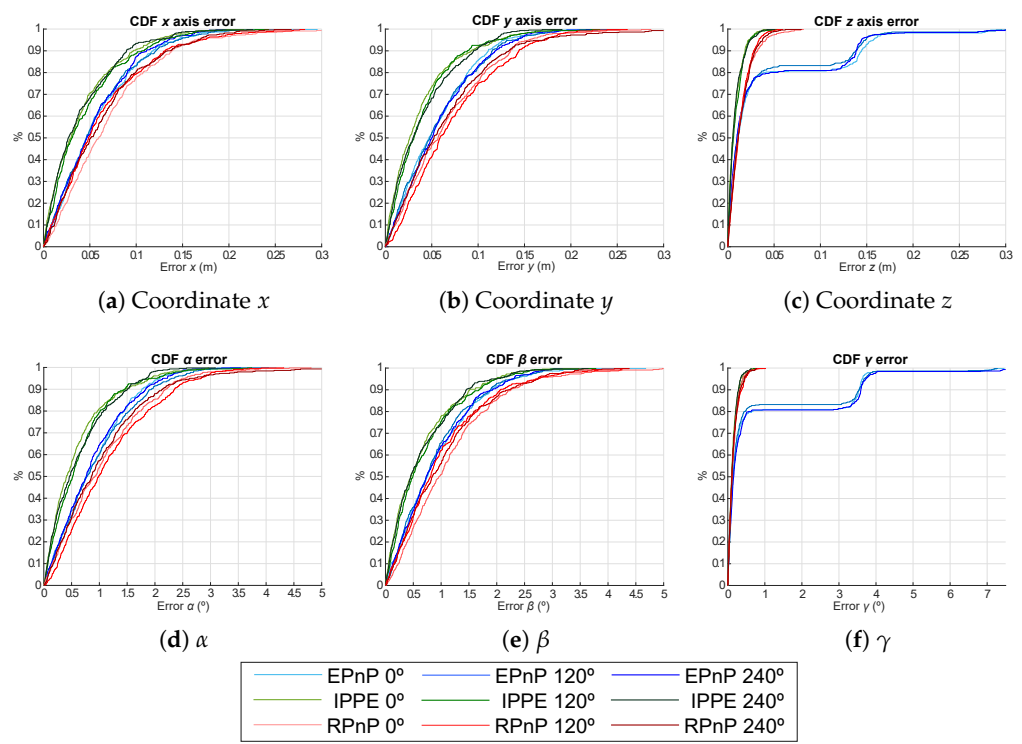

**Figure 10.** CDF of the absolute pose errors for EPnP, IPPE and RPnP algorithms and a rotation in the Z axis $\gamma = \{0°, 120°, 240°\}$ when varying the location of a transmitter with a Gaussian noise $\sigma = 1$ cm.

## 6. Experimental Results

The experimental tests have been conducted in a room of $2 \times 2$ m$^2$ with a height of 3.4 m under normal light and noise conditions (see Figure 11). The LED beacons have been located on the ceiling of the room in its central part distributed in the four corners of a square with a 1.2 m length, whereas the receiver is placed on the floor of the room. It has been verified in simulations that the most suitable algorithm in this proposal is the IPPE algorithm; therefore, hereinafter, only this algorithm has been considered for estimating the pose of the receiver in experimental tests.

The receiver has been calibrated so that all the parameters in (3) are known. These parameters are obtained by minimizing the positioning error by using iterative methods, such as a Branch and Bound algorithm and a Linear Least Squares algorithm. The aperture height is set at $h_{ap} = 2.55$ mm, the optical centre is located at $(c_x, c_y) = (0.055, -0.035)$ mm, the focal length adjustment is $\lambda = 1.25$, the aperture misalignment is $\delta = 0.1$ rad and the aperture length is $l = 2.75$ mm. In addition, an Optitrack system [33] has been applied to obtain the real positions (ground truth) of the transmitting beacons and the receiver. This high-accuracy motion capture system is based on a set of cameras placed around the room that permit obtaining the position of the object under analysis with an accuracy of 0.1 mm. Note that the measurements with the Optitrack system are not simultaneous (they are made sequentially), as the IR emitters used by the Optitrack system saturates the QADA receiver. Three points have been experimentally analysed, as observed in Figure 12 for $\gamma = \{0°, 120°, 210°$ and $300°\}$. These angles are selected to verify that all the quadrants behave similarly.

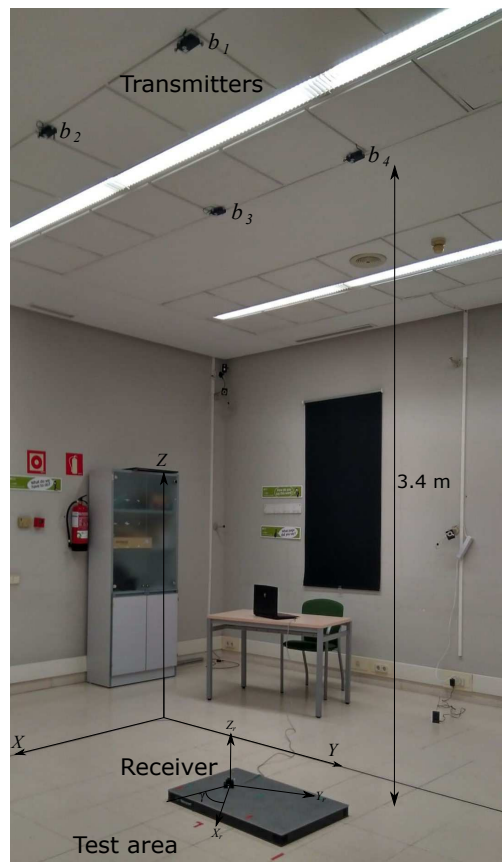

**Figure 11.** Experimental setup in the proposed scenario.

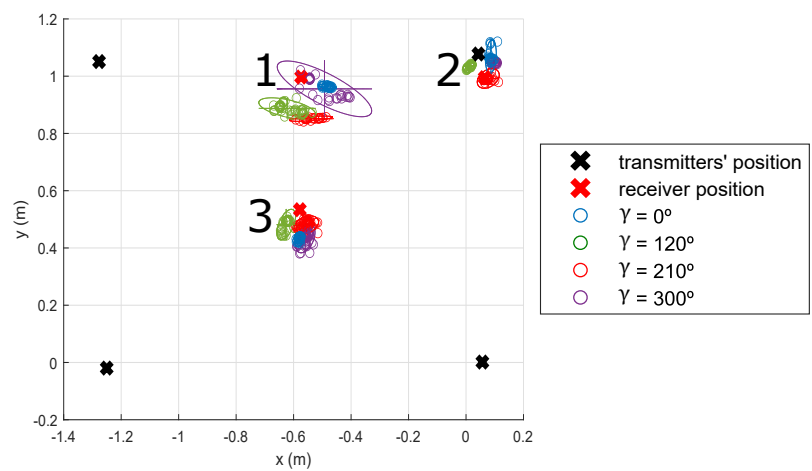

**Figure 12.** Experimental position estimates at $z = 0$ m for $\gamma = \{0°, 120°, 210°$ and $300°\}$ with the IPPE algorithm.

Table 2 details the average and median absolute errors as well as the standard deviation for the analysed points with $\gamma = \{0°, 120°, 210°$ and $300°\}$, performing 50 realizations at each point. The obtained results for coordinates *x*, *y* and *z* are 4.33 cm, 3.51 cm and 28.90 cm for the total average absolute error and 1.84 cm, 1.17 cm and 19.80 cm for the standard deviation, respectively. In addition, the obtained results for the rotation angles $\alpha$, $\beta$ and $\gamma$ are 5.36°, 8.53° and 6.87° for the total average absolute error and 3.07°, 4.22° and 0.63° for the standard deviation, respectively. It is worth mentioning that points beyond $3\sigma$ are treated as outliers and are not taken into consideration in the analysis. Note that $3\sigma$ includes the 99.73 % of all the data.

**Table 2.** Experimental average, median and standard deviation of the absolute error for the three analysed points with $\gamma = \{0°, 120°, 210°$ and $300°\}$ using the IPPE algorithm.

| Points | $\gamma$(°) | Average Absolute Error | | | | | | Median Absolute Error | | | | | | Standard Deviation | | | | | |
|---|---|---|---|---|---|---|---|---|---|---|---|---|---|---|---|---|---|---|---|
| | | X (cm) | Y (cm) | Z (cm) | $\alpha$(°) | $\beta$(°) | $\gamma$(°) | X (cm) | Y (cm) | Z (cm) | $\alpha$(°) | $\beta$(°) | $\gamma$(°) | X (cm) | Y (cm) | Z (cm) | $\alpha$(°) | $\beta$(°) | $\gamma$(°) |
| Point 1 | 0° | 5.59 | 3.17 | 20.65 | 1.41 | 9.55 | 5.54 | 5.53 | 2.86 | 19.09 | 1.03 | 11.07 | 5.57 | 0.47 | 0.77 | 7.48 | 1.22 | 2.63 | 0.15 |
| | 120° | 1.79 | 1.16 | 7.74 | 2.78 | 13.02 | 4.82 | 1.61 | 0.59 | 8.29 | 1.05 | 13.27 | 4.51 | 1.16 | 1.22 | 13.71 | 2.70 | 5.37 | 0.65 |
| | 240° | 7.64 | 0.67 | 27.06 | 3.26 | 18.49 | 5.93 | 8.38 | 0.49 | 31.22 | 1.63 | 20.14 | 5.99 | 1.58 | 1.35 | 13.36 | 3.48 | 4.65 | 0.49 |
| | 300° | 5.80 | 3.86 | 34.91 | 5.22 | 9.52 | 5.32 | 5.61 | 3.45 | 35.59 | 4.86 | 7.80 | 5.20 | 1.42 | 1.43 | 9.43 | 3.27 | 5.45 | 0.42 |
| Point 2 | 0° | 9.50 | 3.14 | 123.29 | 8.73 | 9.64 | 7.57 | 9.92 | 3.15 | 126.22 | 9.46 | 10.30 | 7.55 | 1.07 | 0.31 | 6.55 | 1.89 | 1.62 | 0.18 |
| | 120° | 2.25 | 4.32 | 9.73 | 5.08 | 2.42 | 7.84 | 4.50 | 4.03 | 19.79 | 4.62 | 9.35 | 7.77 | 3.80 | 1.06 | 23.08 | 2.38 | 4.24 | 0.54 |
| | 240° | 3.77 | 6.52 | 22.02 | 9.22 | 3.90 | 7.94 | 2.09 | 6.29 | 31.67 | 9.09 | 3.96 | 7.26 | 2.94 | 0.77 | 25.62 | 4.70 | 3.55 | 1.22 |
| | 300° | 7.05 | 2.14 | 3.92 | 5.62 | 2.32 | 6.47 | 3.44 | 1.18 | 9.81 | 4.08 | 26.30 | 6.55 | 6.31 | 2.70 | 48.32 | 3.83 | 3.21 | 0.97 |
| Point 3 | 0° | 1.90 | 6.84 | 8.54 | 3.63 | 0.44 | 7.48 | 1.94 | 6.07 | 7.48 | 1.66 | 1.36 | 7.36 | 0.62 | 2.06 | 31.86 | 4.24 | 5.17 | 0.75 |
| | 120° | 4.52 | 5.26 | 69.42 | 3.88 | 13.98 | 9.43 | 4.29 | 5.35 | 70.01 | 4.09 | 14.32 | 9.61 | 0.57 | 0.84 | 4.12 | 1.26 | 2.80 | 0.53 |
| | 240° | 0.46 | 0.21 | 2.37 | 6.65 | 9.03 | 8.69 | 0.51 | 0.36 | 1.13 | 6.82 | 11.64 | 8.69 | 1.36 | 1.16 | 27.16 | 3.99 | 6.80 | 0.37 |
| | 300° | 1.66 | 4.88 | 17.16 | 8.82 | 10.11 | 5.39 | 1.13 | 4.86 | 10.06 | 10.58 | 13.48 | 4.49 | 0.81 | 0.39 | 26.92 | 3.92 | 5.13 | 1.29 |
| Total | | 4.33 | 3.51 | 28.90 | 5.36 | 8.53 | 6.87 | 4.08 | 3.22 | 30.86 | 4.92 | 11.92 | 6.71 | 1.84 | 1.17 | 19.80 | 3.07 | 4.22 | 0.63 |

Figure 13 shows the CDF of the absolute errors of the receiver pose estimation. Those points located near the corners of the room (points 1 and 2) present slightly higher errors than point 3, which is located near the centre of the room. Note that the rotation angle $\gamma$ presents a 4 ° absolute error offset that may be due to a slight $\delta$ miscalibration.

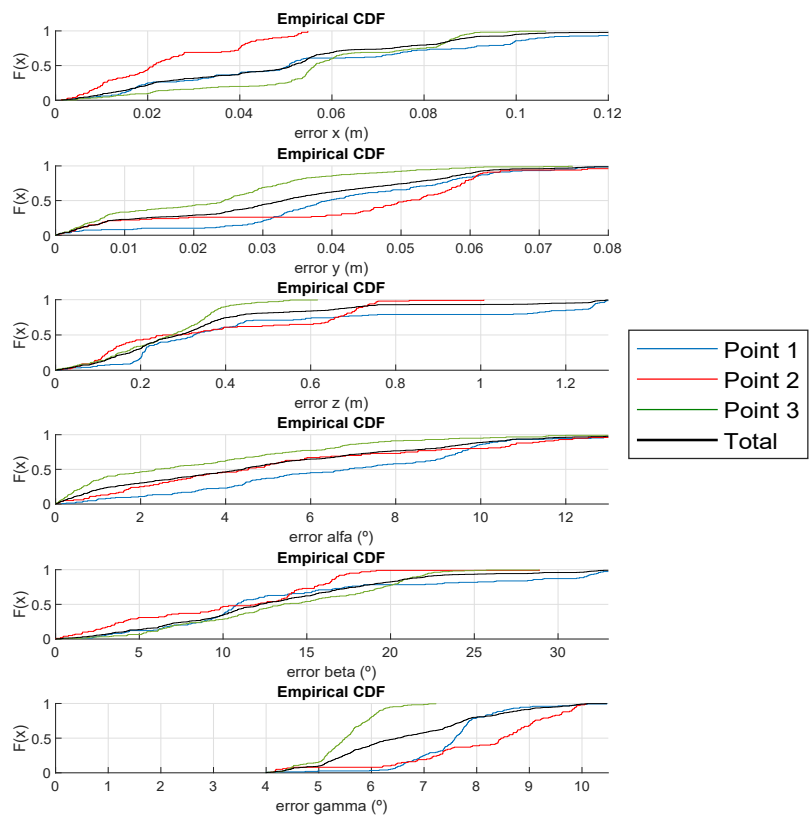

**Figure 13.** CDF of the absolute errors for every experimental point at $z = 0$ m for $\gamma = \{0°, 120°, 210°$ and $300°\}$ with the IPPE algorithm.

A comparison between some previous works considered in the literature is presented in [19]. Several of those works implement different multiplexing techniques to increase their bandwidth and develop Visible Light Communication (VLC) links [34–37], but none of them perform any encoding (although a few implement an OOK modulation). It is worth noting that some involve other sensors to increase their area of operation while decreasing the positioning error (<10 cm), such as mechanical ones (IMU: Intertial Measurement Units) [38] or accelerometers [36]. The investigations [35,36,39] with a small operating area (up to $1 \times 1$ m$^2$) reached positioning errors in the range of 5 cm. On the contrary, in [37], the proposal is tested in a high operating area ($5 \times 5$ m$^2$), resulting in significant positioning errors in the range of 20 cm.

Summing it up, the solution proposed here provides an average positioning error in the scale of previous works with the addition of estimating the entire pose of the receiver ($x$, $y$, $z$, $\alpha$, $\beta$ and $\gamma$) in larger spaces (distance between transmitters and receiver of up to 4.5 m).

## 7. Conclusions

This paper have presented a study of the application of PnP techniques in a long-distance 3D infrared indoor positioning system, which is based on four LED transmitters and a QADA receiver with an aperture. The proposal has been characterized in a room with a height of 3.4 m where the transmitters were installed in the ceiling with a planar configuration in a square with a 1.2 m long side, whereas the receiver remains on the floor. Different perspective-n-point approaches, such as EPnP, IPPE and RPnP, have been analysed in order to obtain the pose of the receiver in simulations. The robustness of the system when varying the 3D location of one transmitter has also been analysed in nine test points of the first quadrant of the grid. In general terms, in the estimation of the 2D coordinates ($x$ and $y$), the absolute average errors are below 10 cm for the IPPE and EPnP algorithms and below 20 cm for the RPnP algorithm. The absolute error related to coordinate $z$ is 3 cm, 7 cm and 30 cm for the IPPE, RPnP and EPnP algorithms, respectively. According to the simulations, the algorithm with a lower error in the pose estimation of the receiver is the IPPE algorithm in this proposal. A real setup has been used for the validation of the system. In particular, the position of the receiver has been estimated for three points in the environment with $\gamma =$ {0°, 120°, 210° and 300°}, performing 50 realizations per point and using the Optitrack system as ground-truth. The experimental tests using the IPPE algorithm achieved average absolute errors of 4.33 cm, 3.51 cm and 28.90 cm and standard deviations of 1.84 cm, 1.17 cm and 19.80 cm in the coordinates $x$, $y$ and $z$, respectively.

**Author Contributions:** Conceptualization, E.A.-E., Á.H. and J.U.; methodology, J.U.; software, E.A.-E. and D.M.; validation, E.A.-E., D.P. and Á.H.; formal analysis, D.P.; investigation, Á.H.; resources, J.U.; data curation, E.A.-E. and D.M.; writing—original draft preparation, E.A.-E. and Á.H.; writing—review and editing, J.U. and D.P.; visualization, E.A.-E. and D.M.; supervision, Á.H. and D.P.; project administration, J.U.; funding acquisition, J.U., D.P. and Á.H. All authors have read and agreed to the published version of the manuscript.

**Funding:** This work has been supported by the Spanish Ministry of Science, Innovation and Universities (POM project, ref. PID2019-105470RA-C33, MICROCEBUS project, ref. RTI2018-095168-BC51 and ATHENA project, ref. PID2020-115995RB-I00), the Community of Madrid (CODEUS project, ref. CM/JIN/2019-043 and PUILPOS project, ref. CM/JIN/2019-038) and the Youth Employment Program (ref. PEJ2018-003459-A).

**Institutional Review Board Statement:** Not applicable.

**Informed Consent Statement:** Not applicable.

**Data Availability Statement:** Not applicable.

**Conflicts of Interest:** The authors declare no conflict of interest.